\documentclass[journal]{IEEEtran}
\ifCLASSINFOpdf
\usepackage[pdftex]{graphicx}
\graphicspath{figures/}
\DeclareGraphicsExtensions{.pdf,.jpeg,.png}
\else
\usepackage[dvips]{graphicx}
\graphicspath{{figures/}}
\DeclareGraphicsExtensions{.eps}
\fi
\usepackage[cmex10]{amsmath}
\usepackage{bm}
\usepackage{hyperref}
\usepackage{mathtools}
\usepackage{helvet}
\usepackage{xcolor}
\usepackage{cite}
\usepackage{amsfonts}
\usepackage{ulem}

\usepackage{url}
\begin{document}
\title{Leveraging Voltage-Controlled Magnetic Anisotropy to Solve Sneak Path Issues in  Crossbar Arrays}

\author{Kezhou~Yang, 
and~Abhronil~Sengupta,~\IEEEmembership{Member,~IEEE}
\thanks{Manuscript received August, 2022.}
\thanks{The authors are with the Department of Materials Science and Engineering, School
of Electrical Engineering and Computer Science, The Pennsylvania State University, University Park,
PA 16802, USA. E-mail: sengupta@psu.edu.

The work was supported in part by the National Science Foundation grants ECCS \#2028213 and CCF \#1955815.}}
\maketitle
\begin{abstract}
\small{In crossbar array structures, which serves as an ``In-Memory" compute engine for Artificial Intelligence hardware, write sneak path problem causes undesired switching of devices that degrades network accuracy. While custom crossbar programming schemes have been proposed, device level innovations leveraging non-linear switching characteristics of the cross-point devices are still under exploration to improve the energy efficiency of the write process. In this work, a spintronic device design based on Magnetic Tunnel Junction (MTJ) exploiting the use of voltage-controlled magnetic anisotropy (VCMA) effect is proposed as a solution to the write sneak path problem. Additionally, insights are provided regarding appropriate operating voltage conditions to preserve the robustness of the magnetization trajectory during switching which is critical for proper switching probability manipulation.}
\end{abstract}

\begin{IEEEkeywords}
Voltage-Controlled Magnetic Anisotropy, Magnetic Tunnel Junction, Spintronic devices, Sneak path current, Crossbar array.
\end{IEEEkeywords}
\vspace{-2mm}
\section{Introduction}

Artificial intelligence (AI) has undergone significant development in the past decade and has been applied in various areas such as speech processing \cite{8887564,purwins2019deep}, video object recognition \cite{Feichtenhofer_2019_ICCV} and financial fraud detection \cite{roy2018deep}, among others. Abstracting the functionality of biological neural networks (NN) as computing models, a singular module consists of synapses (which serve as the memory component) and neurons (which perform the compute role). The fundamental mismatch between such memory embedded compute based architectural models and current von-Neumann based computers results in significant area and energy consumption overhead and limits the performance of AI systems on applications with large problem space complexity. For this reason, research interest on hardware systems based on ``In-Memory" compute that is compatible with NN models at a fundamental architectural level have been growing recently. For a network-level design, crossbar array structure provides a promising pathway towards a resource efficient AI hardware platform \cite{Yang2013,Prezioso2015}. 

\begin{figure*}[!t]
\centering
\includegraphics[width=0.65\linewidth]{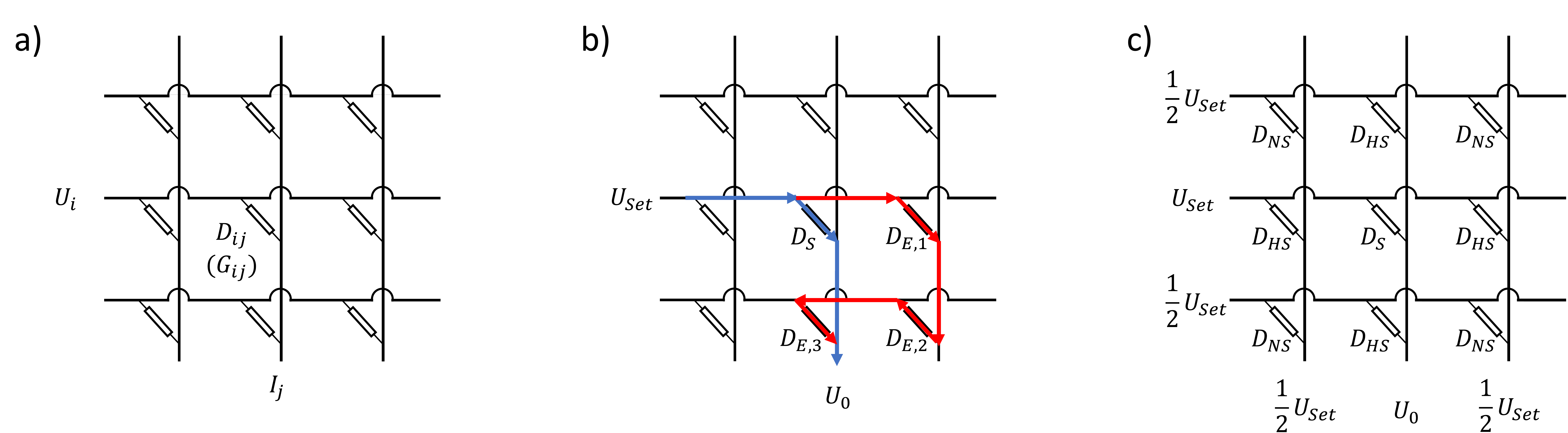}
\caption{a) In a crossbar array structure, voltage signal $U_{i}$ is applied across device $D_{ij}$ with conductance $G_{ij}$ along the horizontal line. The output current signal $I_{j}$ is read from the vertical line. b) Sneak path current can be induced during reading and writing operations of the crossbar array. To read/write a selected cell $D_{S}$, a voltage signal $U_{Set}$ is applied to the device ($U_{0} = 0$), which leads to a read/write current (denoted by blue arrows). On the other hand, this voltage drop also results in sneak path current (denoted by red arrows) passing through other devices ($D_{E,1}$, $D_{E,2}$ and $D_{E,3}$), which causes errors in the reading and writing process. c) Under the custom programming scheme, the selected device $D_{S}$ is under a full set voltage $U_{Set}$. The half-selected devices are under $1/2U_{Set}$ voltage drop. The unselected devices $D_{US}$ are under zero voltage drop. 
}
\label{Fig:Crossbar}
\end{figure*}
In a crossbar array, devices (i.e. synapses) are present at the junction points of the array, as is shown in Fig. \ref{Fig:Crossbar}(a). The crossbar in the figure receives input voltage signals, $U_i$, along the horizontal lines, and produces output current signals, $I_j$, along the vertical lines. Following Kirchhoff's law, output current along the column is given by the equation: $I_j = \sum U_{i}G_{ij}$, where $G_{ij}$ is the conductance of device at the cross-point. Though crossbar array structure is intrinsically efficient in dot-product calculation (key computational primitive required in NN hardware acceleration), errors may occur due to undesired ``sneak path" problems \cite{Sneak1,Shi2020,Cassuto2016}. Sneak path issue refers to the situation where an applied voltage causes undesired current flowing through devices that are not supposed to be read/written, which results in an error in the reading/writing process. For example, in Fig. \ref{Fig:Crossbar}(b), the blue current indicates the desired current which passes through device $D_{S}$, while there may also be the undesired red current along the the other row because there can be a voltage drop across devices $D_{E,1}$, $D_{E,2}$ and $D_{E,3}$ since the other input and output terminals of the array are floating. While sneak path issue occurs both in the reading and writing process, write sneak path issue is a more challenging problem in ``In-Memory" dot-product calculation. 
To solve the sneak path issue, usually a custom programming scheme is adopted for the write process \cite{Chen2003}. As is shown in Fig. \ref{Fig:Crossbar}(c), instead of just applying a voltage signal to terminals linked to the device to switch, all terminals receive a voltage input such that the voltage difference is controlled and sneak path current is mitigated. The device to switch is noted as the ``selected cell" ($D_{S}$) and applied a full set voltage, $U_{Set}$, and the devices noted as ``half-selected cells" ($D_{HS}$) are under a half set voltage, which is not sufficient enough for switching to occur. The remaining cells are ``unselected cells" ($D_{US}$), which experience zero voltage drop. Though the sneak path issue can be reduced under this programming scheme, current flowing through half-selected cells contribute to unwanted energy consumption. For this reason, device-level innovations have been pursued to reduce the energy cost. 

Several device designs based on resistive random access memory (RRAM) have been proposed \cite{Zahoor2020,Ielmini2016,Chang2016}. A typical RRAM device is composed of two metal electrodes and an insulating layer in between. In the insulating layer, a conducting filament (CF) can be formed and modulated when voltage signals are applied across the electrodes. Device conductance is determined by the state of the CF. Previous works have mainly  proposed cell designs consisting of multiple devices, such as 1T-1M \cite{Hu2018,Li2018,Yu2017}, 1D-1M \cite{Song2018,Lee2019} and 1S-1M \cite{Kumar2019,Bae2015} where an additional transistor (T), diode (D) or selector (S) is used to reduce or block the undesired current for half-selected cells. While the energy consumption is reduced due to the mitigation of undesired current, such cross-point designs with multiple devices are not area-efficient and have been succeeded by single-device cell design proposals that leverage intrinsic non-linear $I-V$ characteristics of the cross-point device itself. The cell design exhibiting similar $I-V$ characteristics as a 1S-1M cell is called a self-selective memristor \cite{Huang2017,Woo2013,Son2012} while that behaving  similarly to a 1D-1M cell is called a self-rectifying memristor \cite{Wu2017,Kim2016}. The non-linear $I-V$ characteristics of the single-device cell design enables the reduction of undesired current in a similar manner as the multiple-device cell structure along with a higher area-efficiency. However, such a bit-cell proposal exploiting non-linear $I-V$ characteristics of cross-point devices has not been explored before for spintronic cross-point arrays.

Compared to other non-volatile memory technologies, spintronic devices possesses the advantage of lower operating voltage which reduces energy consumption, faster read and write processes, unlimited endurance and compatibility to conventional CMOS-based systems, which makes it a promising choice to build the next generation hardware platform for neuromorphic computing systems \cite{Verma2020}. However, the intrinsic non-linear physics of emergent novel switching mechanisms of spintronic devices have not been leveraged before to mitigate the sneak path current issue in crossbar array based systems. In this paper, a single-device bit cell solution leveraging voltage-controlled magnetic anisotropy (VCMA) effect is proposed. During the writing process, the full set voltage applied to selected cell switches the device via VCMA effect, while the switching of half-selected cells is still dominated by spin transfer torque (STT). Since the pulse width to achieve high switching probability by VCMA effect is much shorter, the pulse width of set voltage signal can be chosen properly so that a high switching probability is achieved for selected cell while half-selected cells still have a low switching probability tending to zero. The sharp switching probability difference among half-selected cells and selected cells under the applied voltage signal enables a high write accuracy since undesired switching of half-selected cells are restrained. More importantly, the sharp increase in switching probability is due to change in the switching mechanism and is independent of the pulsewidth and can be achieved even with a short pulse in this proposed framework. Compared to STT dominated mechanism, where the sharpness of switching probability increase with pulse amplitude is related to the pulsewidth \cite{garello2018sot}, the proposed framework provides more design-time flexibility. Simultaneously, the proposed solution reduces the system level energy consumption due to short pulse widths required by the VCMA effect for magnetic state switching.

 \section{Preliminaries} \label{Preliminaries}
\begin{figure}[!t]
\centering
\includegraphics[width=\linewidth]{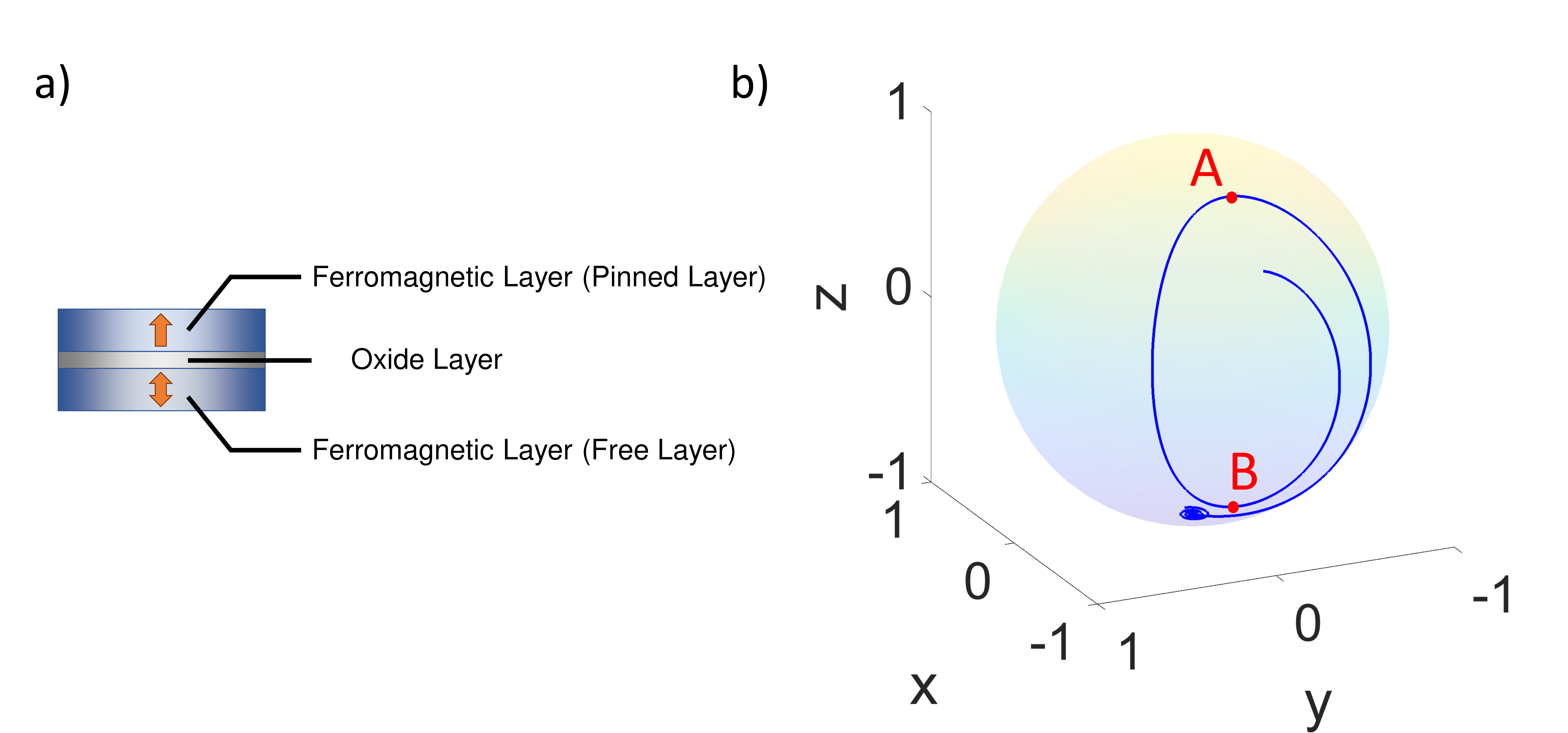}
\caption{a) A typical MTJ consists of two ferromagnetic layers and an oxide layer in between. b) The precession trajectory induced by VCMA effect along in-plane axis is shown. A high switching probability can be achieved if VCMA voltage pulse terminates when magnetization is at point A. The switching probability is low if VCMA pulse terminates when magnetization is at point B.}
\label{Fig:MTJ_Pre}
\end{figure}

\subsection{Device Physics} 

Magnetic tunnel junctions (MTJs) are the basic building block of spintronic devices. A typical MTJ consists of two nanomagnets sandwiching an oxide spacing layer, as is shown in Fig. \ref{Fig:MTJ_Pre}(a). One of the magnetic layers is called ``pinned layer (PL)" since its magnetization is pinned to one direction. The other magnetic layer is called ``free layer (FL)" because its magnetization is free to be switched by external stimuli such as magnetic field or spin current. The state of an MTJ can be defined by the relative configuration of FL and PL magnetization directions. Parallel state (P state) refers to the case where the two layers have the same magnetization direction. P state is associated with a lower electrical resistance while the anti-parallel state with opposite FL and PL magnetic orientation exhibits a higher resistance. The difference in resistance enables information encoding.

For information encoding purposes, spin transfer torque (STT) induced by spin current can be used to switch the device state. But to achieve a high switching probability, a long current pulse is required since the switching probability increases with pulsewidth. To reduce the energy consumption, it is necessary to reduce the pulsewidth of applied pulses. Recent work has shown that short switching pulses can be achieved through voltage-controlled magnetic anisotropy (VCMA) effect \cite{Kanai2014}. VCMA effect enables manipulation of device magnetocrystalline anisotropy energy (MAE, which is the magnetic energy difference between perpendicular and in-plane direction) by an applied voltage via spin-orbit interactions (which consists of two contributions, namely angular momentum and magnetic dipole momentum \cite{Nozaki2019,Suzuki2019}). The orbital angular momentum dominates in strong ferromagnetic materials \cite{Kawabe2017} and can be modulated by doping of charges with selective spin direction. On the other hand, magnetic dipole momentum modification (which results from intra-atomic electron redistribution) is the dominating mechanism in materials where spin-orbit interaction is not strong enough \cite{Miwa2017}. Such VCMA effect enables the change of magnetic anisotropy from perpendicular to in-plane direction.

During the transition of magnetic anisotropy from perpendicular direction to in-plane direction, the FL magnetization performs precession along the in-plane easy axis, as is shown in Fig. \ref{Fig:MTJ_Pre}(b). Considering that the initial magnetization is at the south pole, the switching probability is high when the magnetization stops in the upper half of the unit sphere (region near point A in Fig.  \ref{Fig:MTJ_Pre}(b)) and low in the lower half (region near point B in Fig.  \ref{Fig:MTJ_Pre}(b)), such that the switching probability can be controlled by the pulse width. Prior work has reported that  VCMA-induced switching requires a shorter pulse width that STT-induced switching \cite{Kanai2014}.

The difference in required pulse width between VCMA-induced switching and STT-induced switching leads to a possible solution to the sneak path problem based on the programming scheme illustrated in Fig.  \ref{Fig:Crossbar}(b). If the set voltage is chosen appropriately such that the selected cell operates via VCMA-induced switching  mechanism (although STT is present in this case, the switching process is dominated by the VCMA effect) while half-selected cells operate via STT-induced switching mechanism, the pulse width applied to switch the selected cell will only result in near-zero switching probability to half-selected cells. In this way, a sharp difference in switching probability of selected and half-selected cells can be achieved. 

\subsection{Landau-Liftshiz-Gilbert Equation} \label{LLGsubs}
The behavior of magnetization under an applied external voltage signal that causes STT and VCMA effect can be simulated by Landau-Liftshiz-Gilbert equation \cite{PhysRevB.62.570,PhysRevB.39.6995}.
\begin{equation}
\label{eq:llg}
\frac {d\widehat {\textbf {m}}} {dt} = -\gamma(\widehat {\textbf {m}} \times \textbf {H}_{eff})+ \alpha (\widehat {\textbf {m}} \times \frac {d\widehat {\textbf {m}}} {dt})+\frac{1}{qN_{s}} (\widehat {\textbf {m}} \times \textbf {I}_s \times \widehat {\textbf {m}})
\end{equation}
In Eq. \ref{eq:llg}, $\hat{m}$ is the unit vector in the direction of FL magnetization, $\gamma= \frac {2 \mu _B \mu_0} {\hbar}$ is the gyromagnetic ratio, $\alpha$ is Gilbert's damping ratio,  $N_s=\frac{M_{s}V}{\mu_B}$ is the number of spins in FL of volume $V$ where $\mu_{B}$ is Bohr magneton and $M_{s}$ is saturation magnetization, $q$ is the charge of an electron and $\textbf{I}_{s}$ is the spin current. $\textbf{H}_{eff}$ is the effective magnetic field, including thermal field, demagnetization field and effective magnetic field caused by the VCMA effect. $\textbf{H}_{thermal}=\sqrt{\frac{\alpha}{1+\alpha^{2}}\frac{2K_{B}T_{K}}{\gamma\mu_{0}M_{s}V\delta_{t}}}G_{0,1}$ is used to characterize the thermal noise, where $G_{0,1}$ is a Gaussian distribution with zero mean and unit standard deviation \cite{Scholz2001}. $\textbf{H}_{VCMA}=\frac{2K_{ieff}(U)}{\mu_{0}M_{S}t_{FL}}m_{z}\hat{z}$ is the effective magnetic field caused by the VCMA effect \cite{Sharmin2016}, where $t_{FL}$ is the FL thickness, $m_{z}$ is the $z$ component of $\hat{m}$. $K_{ieff}(U)=K_{i}-\xi\frac{U}{t_{OX}}$ is the expression of effective energy density for interface perpendicular anisotropy, where $K_{i}$ is the energy density of perpendicular anisotropy without applied voltage $U$, $\xi$ is the VCMA co-efficient and $t_{OX}$ is the oxide layer thickness. If not mentioned specifically, simulations are based on parameters mentioned in Table I. The electrical resistance of the MTJ in the P and AP states is obtained from the modelling framework \cite{Sengupta2016} benchmarked to experimental data reported previously in Ref. \cite{Kanai2014}.

\begin{table}[h]
\label{tab:par}
\center
\centerline{TABLE I. Device Simulation Parameters}
\vspace{2mm}
\begin{tabular}{c c}
\hline \hline
\bfseries Parameters & \bfseries Value\\ 
\hline
Free-layer width, $W_{MTJ}$ &  40 nm \\
    Free-layer length, $L_{MTJ}$ & 70 nm \\
    Free-layer thickness, $t_{MTJ}$ & 0.9 nm \\
    Oxide layer thickness, $t_{OX}$ & 1.3 nm \\
    Saturation magnetization, $M_S$ & 1257.3 kA/m \cite{Sengupta2016}\\
    Gilbert-damping factor, $\alpha$ & 0.075 \\
    Temperature, $T$ & 300 K \\
    VCMA co-efficient, $\xi$ & 200 fJ/V$\cdot$ m \\
    Interfacial perpendicular anisotropy, $K_{i}$ & 0.9267 mJ/m$^2$ \cite{Sengupta2016}\\
\hline \hline
\end{tabular}\\ 
\end{table}


\section{Proposal}
\begin{figure*}[t!]
\centering
\includegraphics[width=0.65\textwidth]{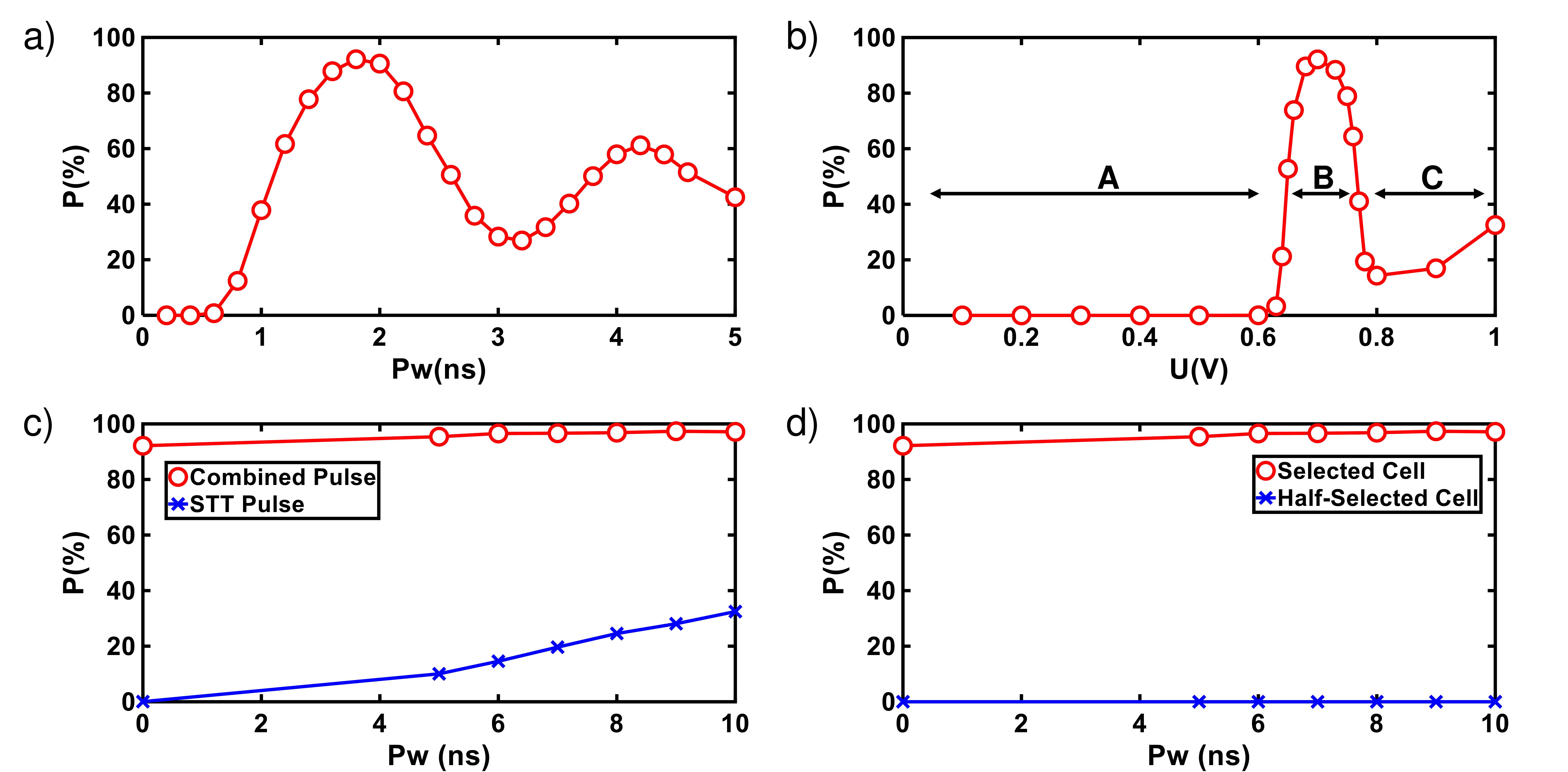}
\caption{a) Switching probability ($P$) changes with applied pulse width ($P_w$) for pulse magnitude $U = 0.7$ V. The precession of device magnetization results in the peaks and valleys. The highest peak depicts a switching probability of $92.1\%$ for a pulse width of $1.8$ ns. b) Variation of switching probability with pulse amplitude is shown. The pulse width is fixed to be $1.8$ ns. c) Switching probability - pulse width (following STT pulse) variation for combined pulsing scheme is given by the red curve and that of a pure STT pulse is given by the blue curve.  d) Variation of selected cell (under $U_{Set}$ voltage corresponding to VCMA-STT pulsing scheme) switching probability  with the following STT pulse width is given by the red curve. Half-selected cell (under $1/2 U_{Set}$ voltage corresponding to VCMA-STT pulsing scheme) switching probability variation with following STT pulse width is given by the blue curve. $U_{Set}$ is a VCMA-STT combined pulse ($0.7$ V, $1.8$ ns VCMA pulse followed by $0.6$ V STT pulse).}
\label{Fig:P_PW}
\end{figure*}
\subsection{Simulation Results}

Fig. \ref{Fig:P_PW}(a) shows the relation between MTJ switching probability and pulse width for $0.7$ V voltage pulses. The fluctuation in switching probability results from the precession along the in-plane axis. The peaks (valleys) are according to the cases where the voltage pulse terminates when the magnetization rotates to the top (bottom) positions of the trajectory. The peaks and valleys tend to $50\%$ switching probability as the magnetization gradually rotates to the in-plane direction with increasing pulse width, which is the new easy axis under VCMA effect. The switching probability is stable at $50\%$ when the FL magnetization remains at the in-plane direction. The $92.1\%$ switching probability at the first peak implies the viability to switch the device with a set voltage pulse as short as $1.8$ ns. Fig. \ref{Fig:P_PW}(b) shows the relation between switching probability and pulse amplitude for $1.8$ ns wide switching pulses. The switching probability remains low for pulses with small amplitude (region A). The reason is that the VCMA effect induced by such low amplitude pulses is not strong enough. In this region, the STT dominates the switching event, which requires a longer pulse for high switching probability. The sharp increase of switching probability at $0.65$ V indicates that the VCMA effect induced by $0.65$ V pulse is strong enough to change the magnetic anisotropy from perpendicular direction to in-plane direction. The $1.8$ ns pulse width enables the magnetization to reach the top half of the trajectory (see Fig. \ref{Fig:MTJ_Pre}(b)), resulting in a switching probability of $92.1\%$ (region B). The switching probability again drops when the pulse amplitude is larger than $0.8$ V (region C). This can be explained by the magnetization trajectory robustness, which will be discussed in the next section.

To further increase the switching probability, another STT pulse can be applied after the VCMA pulse, as is proposed in prior literature \cite{Kanai2014}. In our work, the VCMA-STT combined pulse consists of a short VCMA pulse ($0.7$ V, $1.8$ ns) and a following STT pulse ($0.6$ V). The pulse width of the STT pulse can be fixed in accordance to the desired switching probability of the selected cell. In order to justify the contribution of VCMA switching to the proposal, we also compare against a pure STT pulsing scheme consisting of a first STT pulse ($0.6$ V, $1.8$ ns) and a following STT pulse ($0.6$ V). Fig. \ref{Fig:P_PW}(c) shows that the switching probability of the selected cell under VCMA-STT combined switching increases with the pulse width of the following STT pulse. The switching probability reaches $97\%$ when the following STT pulse is longer than $9$ ns. On the other hand, the switching probability for the pure STT pulse scenario is much lower than that of VCMA-STT combined pulse, which indicates that the VCMA-STT combined pulse can be much shorter (and therefore much more energy efficient) than pure STT pulse to reach a high switching probability for the selected cell.

On the other hand, in the programming scheme shown in Fig. \ref{Fig:Crossbar}(b), there are also half selected cells that experience half set voltage during the switching process. To avoid undesired switching, such half-selected devices should exhibit near-zero switching probability. As is shown in Fig. \ref{Fig:P_PW}(d), the switching probability remains near-zero for the half-selected cell even when the selected cell experiences a switching probability over $97\%$. The sharp difference in switching probability resulting from the VCMA effect makes it possible to ensure a high switching probability for selected cells and a near-zero switching probability for half-selected cells and therefore provides a solution to the sneak path problem for MTJ-based spintronic crosspoint arrays.

\begin{figure}[!t]
\centering
\includegraphics[width=0.65\linewidth]{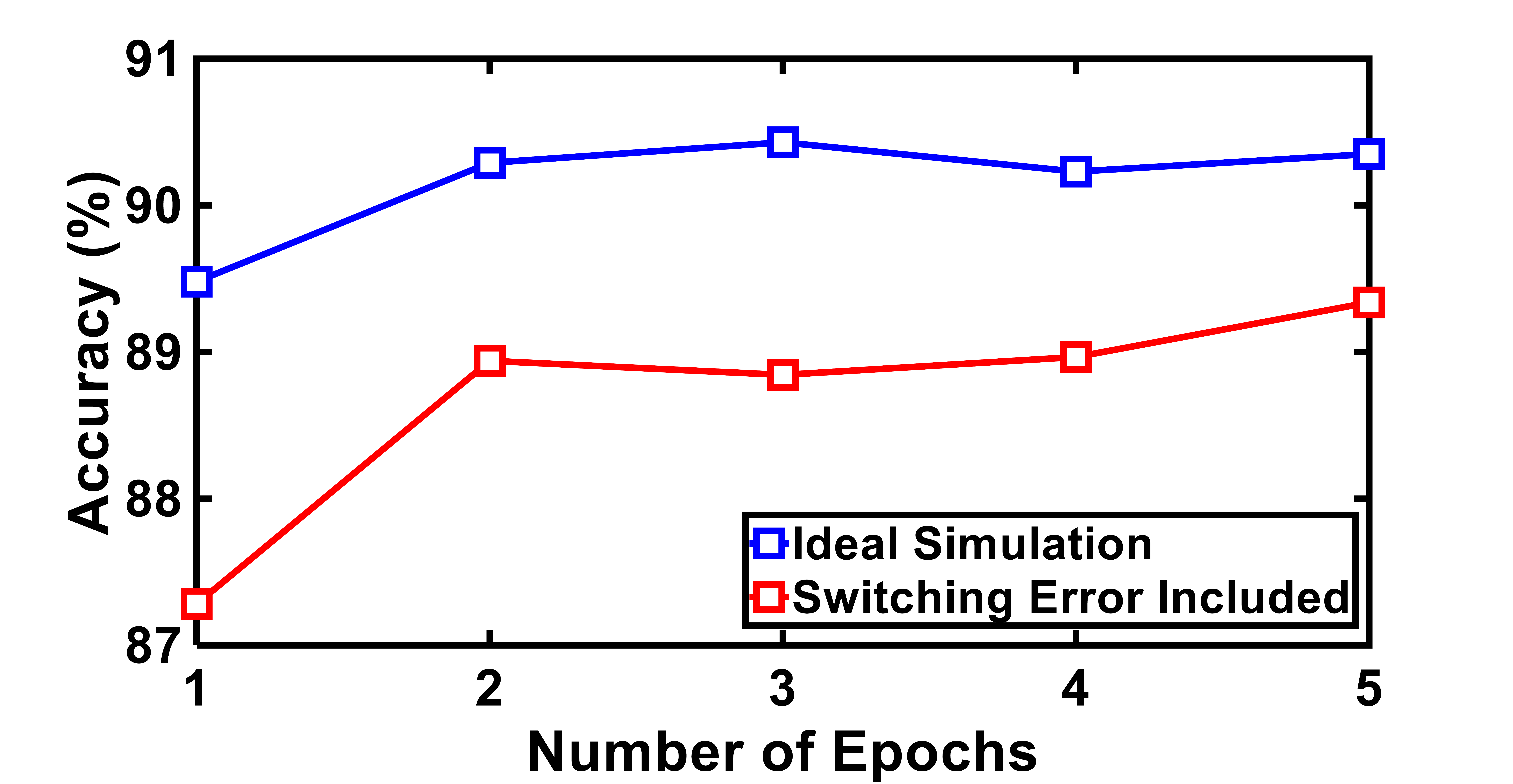}
\caption{Accuracy of network with and without switching error has been obtained for different training epochs. Switching error only causes a reduction of $1.01\%$ in accuracy after 5 epochs of training.}
\label{Fig:CompNw}
\end{figure}

Unlike logic applications, neuromorphic computing applications are resilient to minor imprecision in hardware operation. While the maximum switching probability shown for AP to P switching was $\sim97\%$ (note that P to AP switching will have a slightly reduced switching probability since VCMA pulse is always of the same polarity and STT pulse varies in polarity in the two cases), this did not have any significant impact at the system level. On-chip learning simulations were performed for a $784\times10$ network on the MNIST dataset. The ideal software accuracy was evaluated to be $91.13\%$ ($5$ epochs). The weight values in the network were implemented using $10$-bit resolution. The weight discretized network (considering $100\%$ switching probability in the devices) had an accuracy of $90.35\%$ ($5$ epochs) while the hardware-realistic simulation with slightly reduced switching probabilities had an accuracy of $89.34\%$ ($5$ epochs, averaged for 5 independent runs of the training process.), which is only $\sim1.01\%$ lower than the network with no switching error. The training convergence time is not affected due to the hardware non-idealities and constraints (see Fig. \ref{Fig:CompNw}).

\subsection{Robustness} 

It is observed that even when the magnetization motion is dominated by VCMA effect, switching probability may still be low, as is shown in Fig. \ref{Fig:P_PW}(b) region C. This is due to the loss of magnetization trajectory robustness. The robustness refers to the uniqueness of the route of magnetization precession. The high switching probability results from the fact that every time when the voltage pulse ends, the magnetization is right at the top of the trajectory, which only happens when the magnetization follows the same trajectory. If the magnetization precession trajectory is random, there is no determined relation between pulse width and final position of magnetization. 
In other words, a high switching probability can be ensured only when there is a certain magnetization trajectory (i.e., the robustness is preserved).
\begin{figure}[t]
\centering
\includegraphics[width=0.53\textwidth]{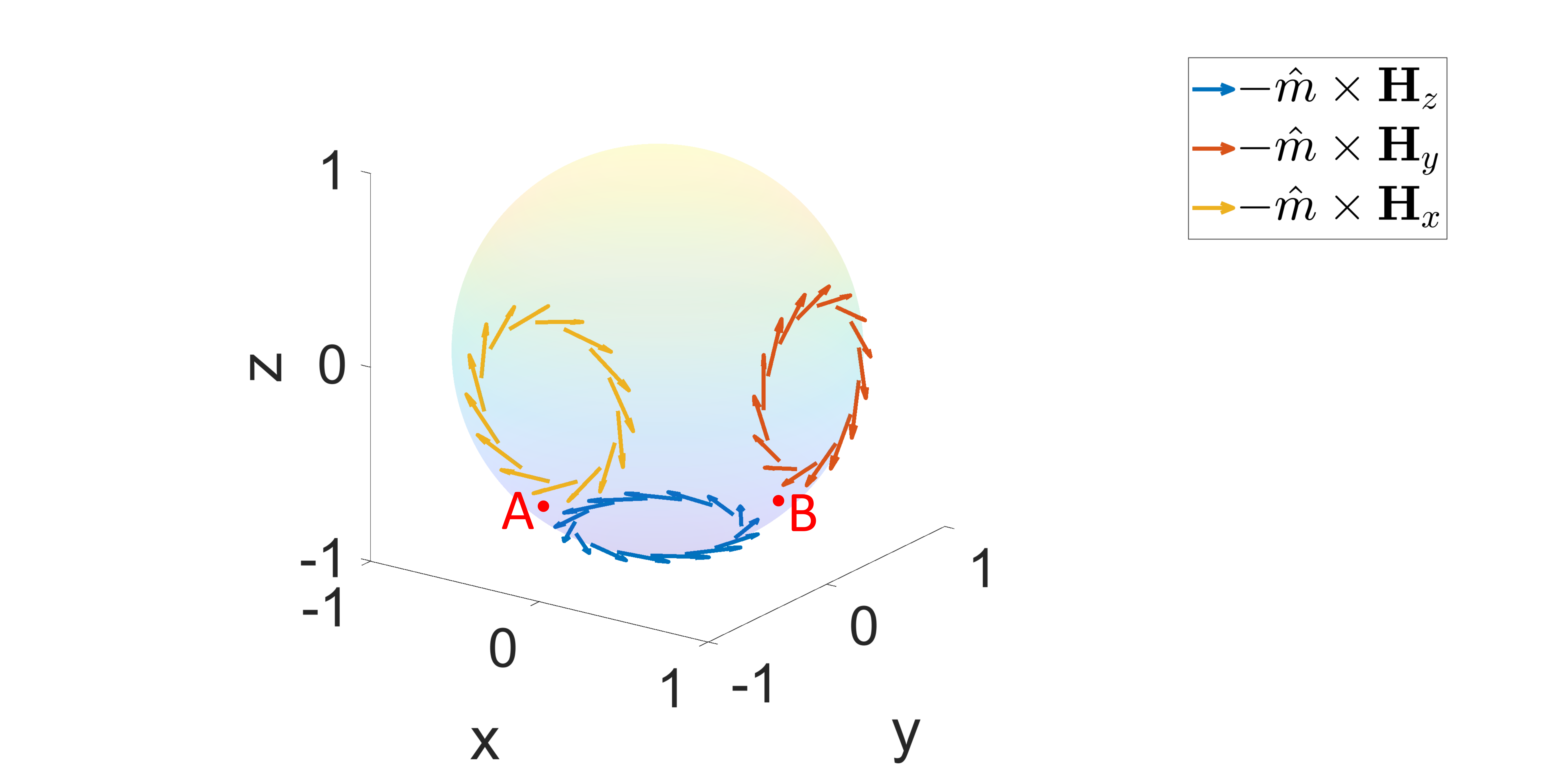}
\caption{The field vectors $-\hat{m} \times \textbf {H}_{x}$, $-\hat{m} \times \textbf {H}_{y}$ and $-\hat{m} \times \textbf {H}_{z}$ under $U = 0.7$ V are plotted on the unit sphere. Each of the fields lead to a precession of the magnetization along the corresponding axis, with a repelling component. Since the field vectors have components along opposite directions in the adjacent region between any two pairs of the three fields, the direction of the total field depends on the relative magnitude  of $\textbf{H}_{x}$, $\textbf{H}_{y}$ and $\textbf{H}_{z}$.}
\label{Fig:dmdt0.7V}
\end{figure}

\begin{figure}[t]
\centering
\includegraphics[width=0.5\textwidth]{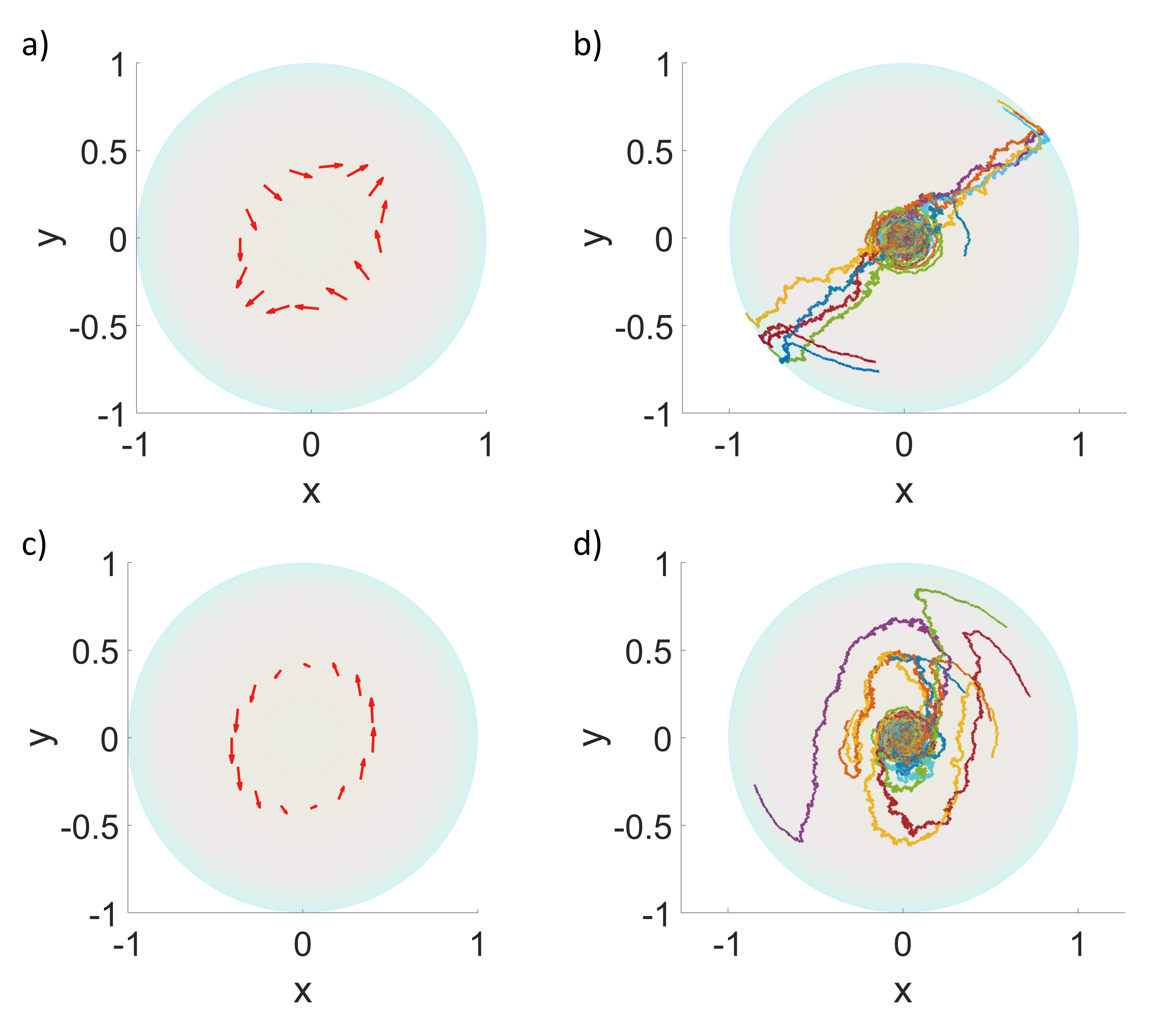}
\caption{a) Total $d\hat{m}/dt$ field under applied voltage $U = 0.7$ V in the region around the south pole of the unit sphere. The relative magnitude of $\textbf{H}_{x}$, $\textbf{H}_{y}$ and $\textbf{H}_{z}$ results in two symmetric exit windows along diagonal directions in the $X-Y$ plane. b) Trajectories of 10 LLG simulations leave the pole area from the exit windows, which enables stable magnetization motion. c) Total $d\hat{m}/dt$ field under applied voltage $U = 0.8$ V in the region around the south pole. In this situation, $\textbf{H}_{z}$ dominates and the total field does not form definite exit windows unlike the $U = 0.7$ V case. d) Trajectories of 10 LLG simulations for applied voltage $U = 0.8$ V are almost random.}
\label{Fig:Rob}
\end{figure}

In order to figure out how the robustness is preserved, it is necessary to study the motion of FL magnetization, $\hat{m}$. The FL magnetization motion can be characterized by the motion ``velocity" on the unit sphere, $d\hat{m}/dt$, which is given by the LLG equation in Eq. \ref{eq:llg}. The precession is mainly related to the first term, $-\hat{m} \times \textbf {H}_{eff}$, where VCMA effect contributes by adding an effective field $\textbf{H}_{VCMA}$ in $\hat{z}$ direction. Denoting the total magnetic field along $\hat{x}$ (short axis), $\hat{y}$ (long axis) and $\hat{z}$ (perpendicular axis) directions as $\textbf{H}_{x}$, $\textbf{H}_{y}$ and $\textbf{H}_{z}$ respectively, Fig.  \ref{Fig:dmdt0.7V} shows the direction of the vector field $-\hat{m} \times \textbf {H}_{x}$, $-\hat{m} \times \textbf {H}_{y}$ and $-\hat{m} \times \textbf {H}_{z}$ under $V = 0.7$ V. Any component of $\textbf{H}_{i} (i \in {x,y,z})$ forms a precession along axis $i$ solely. Note that since $\hat{m} \cdot \textbf {H}_{i} < 0$, there is also a component repelling $\hat{m}$ from axis $i$. 
Direction of the total field depends on the magnitude relation among $\textbf{H}_{x}$, $\textbf{H}_{y}$ and $\textbf{H}_{z}$. For example, at the position of point A, where $\textbf{H}_{y}$ is small and can be neglected, total field direction will be in the same direction as $\textbf{H}_{x}$ ($\textbf{H}_{z}$) if $|\textbf{H}_{x}| > |\textbf{H}_{z}|$ ($|\textbf{H}_{x}| < |\textbf{H}_{z}|$). For the same reason, total field at point B is dominated by the larger component between $\textbf{H}_{y}$ and $\textbf{H}_{z}$. As a result, competition among $|\textbf{H}_{x}|$, $|\textbf{H}_{y}|$ and $|\textbf{H}_{z}|$ leads to a different total field.

Fig. \ref{Fig:Rob}(a) shows the total $d\hat{m}/dt$ field for applied voltage $U = 0.7$ V around the south pole, which is the initial location of the device magnetization. In this situation, the magnitude of the magnetic field satisfies $|\textbf{H}_{x}| < |\textbf{H}_{z}| < |\textbf{H}_{y}|$, which results in two symmetric exit windows. Magnetization trajectory robustness can be preserved since the precession starts from one of the two symmetric exit windows every time. Fig. \ref{Fig:Rob}(b) shows the magnetization trajectories of 10 LLG runs for $U = 0.7$ V. Magnetization leaves the north pole from one of the two exits every time and follows a certain trajectory. In this case, the switching probability can be controlled by the applied pulse width since the relation between the position of magnetization and pulse width is determined by the fixed trajectory. On the other hand, for larger applied voltage, the magnitude of magnetic field satisfies $|\textbf{H}_{x}| < |\textbf{H}_{y}| < |\textbf{H}_{z}|$. Fig. \ref{Fig:Rob}(c) shows the total $d\hat{m}/dt$ field for applied voltage $V = 0.8$ V. There is no definite exit-points in the pole region in this situation, as is shown in Fig. \ref{Fig:Rob}(d). The trajectories are random and the switching probability cannot be controlled by the applied pulse width. 
On the other hand, although the magnetization trajectories are random in this case, the switching probability is still increasing with pulse amplitude, as is shown in Fig. \ref{Fig:P_PW}(b). The reason is that the magnetic energy in $z$ direction increases with applied pulse amplitude due to increasing $|\textbf{H}_{z}|$ caused by the VCMA effect. Due to larger magnetic energy, for larger pulse amplitude, the magnetization is more likely to leave the pole and switch to the other side when the pulse ends, leading to a higher switching probability.

As a result, to enable a high switching probability, the applied voltage $U$ has to be in a range determined by device parameters:
\begin{equation}
\begin{aligned}
\label{eq:Condition}
\frac{t_{ox}}{\xi}(\frac {(N_{xx}-N_{zz})M_S^2\mu_0t_{FL}}{2}+K_i) <\\
U<\frac{t_{ox}}{\xi}(\frac {(N_{yy}-N_{zz})M_S^2\mu_0t_{FL}}{2}+K_i)
\end{aligned}
\end{equation}
where, $N_{xx}$, $N_{yy}$ and $N_{zz}$ are demagnetization factors determined by the device shape. Other parameters are the same as the ones introduced in Section \ref{LLGsubs}.  Since VCMA is a surface
effect occurring at the interface between the free layer and the
oxide layer, variations in free layer thickness play a critical
role in ensuring that a given operating voltage range is robust
enough to device variations. The robustness of the operating voltage range to device parameter variations implies that there should be an overlapping region in the operating voltage ranges of all devices in the system such that they can be programmed at the same voltage. To verify the operating voltage range robustness, the operating window of 1000 devices was calculated according to Eq. 2. and $6\sigma = 1.5\%$ \cite{Sharmin2016} variation in the free layer thickness was considered. It was found that over $99\%$ of all 1000 devices can work at the same applied voltage in the operating voltage range between $0.68V$ to $0.73V$. It is worth mentioning here that other device, circuit and system level parameters like Gilbert's damping ratio, input pulse shape waveform variations can also influence the magnetization reversal in the time domain \cite{ikegami2017voltage}.

\section{Conclusions}

In this paper, a spintronic device utilizing VCMA effect induced switching scheme is proposed as a solution to the sneak path problem in neuromorphic non-volatile cross-point arrays. The required pulse width difference between VCMA induced switching and spin transfer torque switching leads to a sharp difference in switching probability (over $97\%$ for VCMA induced switching and $\sim 0\%$ for STT switching) and thereby enables a potentially energy efficient solution to the write sneak path problem. Additionally, it is also observed that ensuring a specific operating voltage range is critical for the VCMA effect to ensure high switching probability of selected cells such that the effective magnetic field $\textbf {H}_{z}$ does not exceed $\textbf {H}_{x}$, which leads to the loss of FL magnetization trajectory robustness.

\vspace{-2mm}


\end{document}